\documentclass[preprint,showpacs,preprintnumbers,amsmath,amssymb,nofootinbib]{revtex4-1}

\usepackage{graphicx}
\usepackage{amsmath}
\usepackage{amssymb}
\usepackage{xcolor}
\usepackage{dcolumn}
\usepackage{epstopdf}

\begin{document}

\title{Search for strong gravitational lensing effect \\in the current GRB data of BATSE}

\renewcommand{\thefootnote}{\fnsymbol{footnote}}

\author{Chun-Yu Li$^{1*}$ and Li-Xin Li$^{2}$}
\affiliation{$^{1}$Department of Astronomy, Peking University,\\ Beijing 100871, China;\\
$^{2}$Kavli Institute for Astronomy and Astrophysics, Peking University,\\ Beijing 100871, China}

\footnotetext{Corresponding author(email: licy08@pku.edu.cn)}

\date{\today}

\begin{abstract}
Because gamma-ray bursts (GRBs) trace the high-z Universe, there is an appreciable probability for a GRB to be gravitational lensed by galaxies in the universe. Herein we consider the gravitational lensing effect of GRBs contributed by the dark matter halos in galaxies. Assuming that all halos have the singular isothermal sphere (SIS) mass profile in the mass range $10^{10} h^{-1} M_\odot < M < 2\times 10^{13} h^{-1}M_\odot $ and all GRB samples follow the intrinsic redshift distribution and luminosity function derived from the \emph{Swift} LGRBs sample, we calculated the gravitational lensing probability in BATSE, \emph{Swift}/BAT and \emph{Fermi}/GBM GRBs, respectively. With an derived probability result in BATSE GRBs, we searched for lensed GRB pairs in the BATSE 5B GRB Spectral catalog. The search did not find any convincing gravitationally lensed events. We discuss our result and future observations for GRB lensing observation.

\vspace{0.3cm}

\noindent
{\bf  Star formation, Galactic halos, Gravitational lenses and luminous arcs, $\gamma$-ray sources; $\gamma$-ray bursts, Cosmology}

\vspace{0.3cm}

\end{abstract}

\maketitle

\section{Introduction}
After about 50 years since its discovery, the gamma-ray burst (GRB) phenomenon is still one of the most researched topics in modern astrophysics. In recent years, observations have advanced greatly with space instrumentations including burst and transient sources experiment (BATSE), \emph{Swift} and \emph{Fermi}. Since GRBs have high redshifts, they can be used to probe the early universe. For high redshift sources like quasars and GRBs, the gravitional lensing probability is significant and researchers expect gravitational lensing phenomenon will be detected. It is known that gravitational lensing of quasars has already been observed for decades, but the lensing of GRB has never been detected to date.

 Compared to quasars, GRBs are brighter and their average redshifts are higher, so gravitational lensing of GRBs would be considered easier. Dozens of gravitationally lensed quasars have been identified in tens of thousand quasars \citep{ina12}, but no gravitational lensed GRB has ever been identified to date. Nemiroff et al. \cite{nem01} attempted to search for millilensing (with time delay in dozens of second) in $774$ GRBs of BATSE, Ougolnikov \cite{oug01} also reported trying to find  Mesolensing (due to globular clusters with mass about $10^6M_{\odot}$) in $1,512$ BATSE GRBs sample, and Komberg et al. \citep{kom99} attempted to look for the lensing of GRBs in the BATSE 4B Catalog. However, these researchers did not identify any lensing case. Since the sample size has increased from $1,637$ to $2,702$ in the BATSE Current GRB Catalog (http://www.batse.msfc.nasa.gov/batse/grb/catalog/current/), and $2,145$ of them have been analysed in the BATSE 5B GRB Spectral Catalog \cite{ada13}, herein we try to search for the gravitational lensing event of GRBs in the new spectral catalog.

 We calculate the lensing probability in BATSE, \emph{Swift}/BAT and \emph{Fermi}/GBM GRB samples and estimate the expected period to observe one lensing GRB pair for each detector. Later, we search for GRB gravitational lensing pairs in the BATSE 5B GRB Spectral Catalog.
\section{Gravitational Lensing Probability in GRB Samples by Dark Matter Halos in Galaxies}
 As reported elsewhere \cite{li02,li03,li14}, herein we consider the gravitational lensing effect contributed by the dark matter halos in galaxies. The mass function of the halos is given by the Press-Schechter function and the Universe is described by the standard LCDM model. We assume that all halos have the singular isothermal sphere (SIS) mass profile in the mass range $10^{10} h^{-1} M_\odot < M < 2\times 10^{13} h^{-1}M_\odot $. Halos with masses out of this range often have a NFW-type mass profile and make smaller contribution to the total lensing probability \citep{li02,li03}.

Lensing probability in normal source samples like quasars and galaxies can be calculated with integrated lensing probability $P(z)$, the redshift probability density distribution for a random sample, or the normalized redshift distribution of the complete sample $n(z)$ and the lensing magnification bias $B(z)$. The expected sample size to observe one lensing case is given by \cite{li14}
\begin{eqnarray}
	N_0=1/P= \frac{1}{\int^\infty_{z=0}\ B(z) P(z)n(z)dz}. \label{N0}
\end{eqnarray}

Note that this equation indicates in every $N_0$ samples there is probably one sample which has been lensed by halos. For galaxies and quasars, if the sample is lensed with proper angular separation and brightness ratio, we can observe the double image at the same time, thereby identifying the lensing case. However, for a transient source like GRB, because of  the short duration compared to the lensing time delay, researchers generally do not catch the lensed double burst at the same time. Since usually we are not able to keep observing at the same field for a long time, it may be possible missing to identify the lensing case even if it does occur. Thus the expected sample size should be larger than $N_0$ because of observation constraints. We quantitatively define the observation condition influence by the sampling efficiency .

In Eq. (\ref{N0}), the lensing probability $P(z)$ for a remote point source at redshift $z$ with brightness ratio $r \leq 5$ is calculated in \cite{li14}. Here we calculate the magnification bias $B(z)$, the sample's redshift distribution $n(z)$, and sampling efficiency $f$.
\subsection{Reconstructing the Redshift Distribution of \emph{Swift}/BAT LGRBs Sample}
\begin{figure}
\begin{center}\includegraphics[angle=0,scale=0.65]{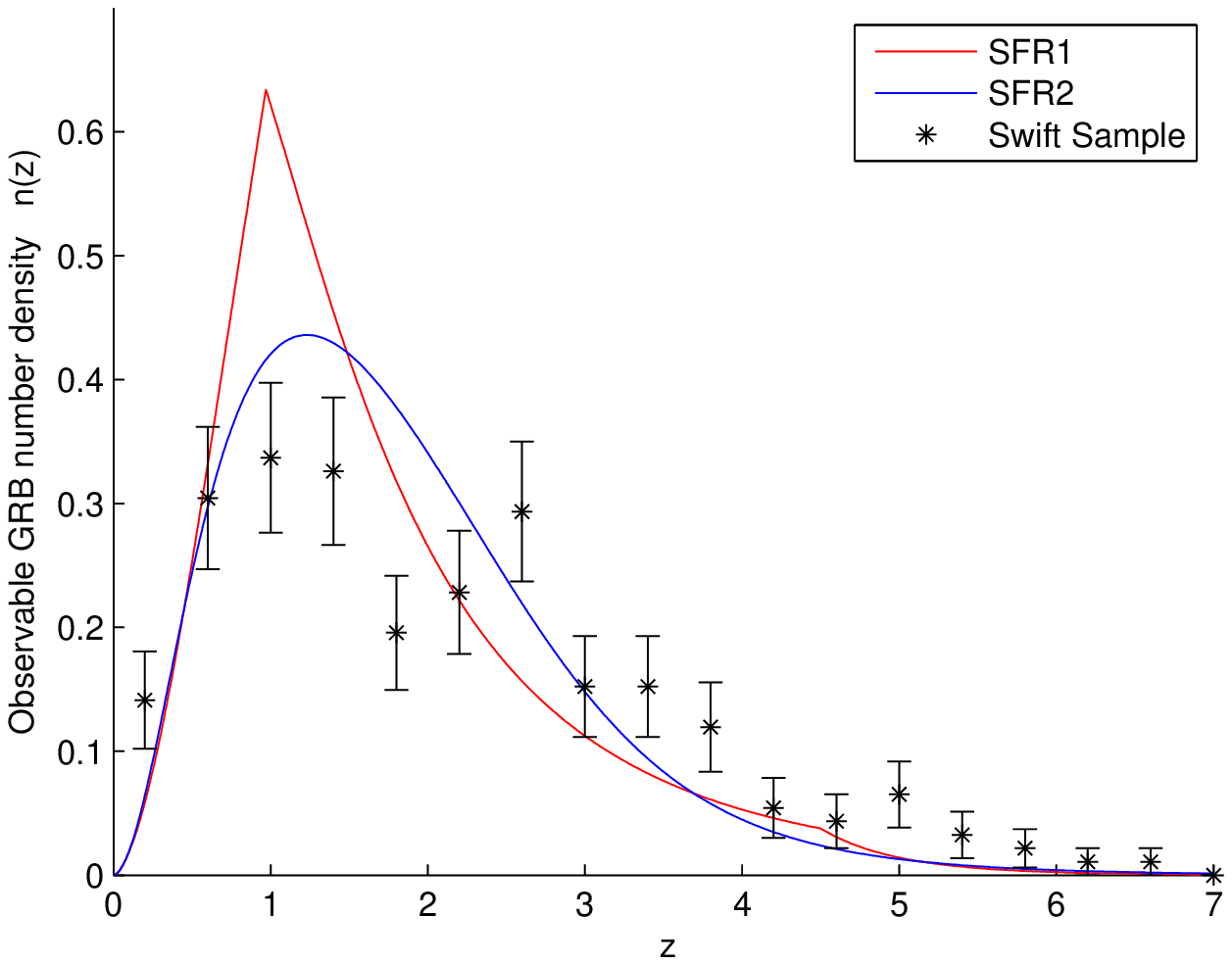}\end{center}
\caption{The normalized observable GRB number distribution in the redshift space for \emph{Swift}. The red line uses the SFR in
 Eq. (\ref{SFR1}), and the blue line uses the SFR in Eq. (\ref{SFR2}). Points with error bars are the samples of 233 \emph{Swift} LGRBs up to GRB140323A with redshift measurement. The bin width in redshift is $0.4$.
\label{fig1}}
\end{figure}

Swift contributes a large fraction of GRBs with redshift measurement. Up to GRB140323A, there are $233$ LGRBs with redshift data in Swift's total $743$ sample. In  Fig. \ref{fig1}, we show the normalized redshift distribution of the $233$ samples. The discrete points with error bar are sample number with bin width $z=0.4$. We can see a readily apparent sample missing in $z=[1,\ 2.5]$, because of the redshift dersert effect. But this effect can not explain the large fraction ($2/3$) redshfit lost of the whole sample. Thus directly using the incomplete sample data to get the redshift distribution will cause large uncertainty.

Most researchers have tried to reconstruct the redshift distribution and luminosity function of GRBs based on \emph{Swift} LGRB sample \citep{li08,cao11,tan13}. Assuming the GRBs distribution proportionally following the star formation rate (SFR) history and considering some evolution factor such as beam factor and metallicity, the redshift distribution and luminosity function for a given model can be fit. In this paper, we use the redshift distriubtion and luminosity function fitted in \cite{tan13}, calling the model SFR1. To make comparison, we also use another SFR fitting formula, terming the model SFR2.

The luminosity function of GRB is fitted with \emph{Swift} LGRB given by
\begin{eqnarray}
\Phi(L)\propto\left\{~\begin{array}{ll}(\frac{L}
{L_b})^{-v_1},~~~~& L\leq L_b,\,\\
(\frac{L}
{L_b})^{-v_2},~~~~&L>L_b,\,\end{array}\right.
\label{lum}
\end{eqnarray}
where $v_1=1.2$, $v_2=2$, $L_b = 1.3 \times10^{52}\ erg\cdot s^{-1}$.

The GRB number density in redshift space is given by \cite{tan13}
\begin{eqnarray}
	\Sigma_{grb}(z) =f_BC \frac{\rho(z)}{1+z} \frac{dV_{com}}{d z},\label{GRB rate}		
\end{eqnarray}
where $f_B$ is the beaming degree of GRB outflows and the proportional coefficient C can arise from the particularities of
GRB progenitors (for instance, mass, metallicity, magnetic field, etc). $f_BC\ =\ 2.4\times10^{-8} (1+z)^{-1} M_{\odot}^{-1}$. $\rho(z)$ is the comoving star formation rate density fitted by \cite{hop06}
\begin{eqnarray}
\rho(z)\propto \left\{~~\begin{array}{ll}(1+z)^{3.44},&
~~~~z\leq0.97,\,\\
(1+z)^{-0.26}, & ~~~~0.97<z\leq4.5,\,\\
(1+z)^{-7.8},& ~~~~z>4.5,\,
\end{array}\right.\label{SFR1}
\end{eqnarray}
where $\rho(0)=0.02~M_{\odot}\rm yr^{-1} Mpc^{-3}$.

For comparison, we also use another SFR fitting form as in \citep{col01}
\begin{eqnarray}
	\rho_0(z) = \frac{a+b z}{1+(z/c)^d} \;, \label{SFR2}
\end{eqnarray}
where $(a,b,c,d) = (0.0157, 0.118, 3.23, 4.66)$  \citep{li08}.

Then, the number of observable GRB between $[z, z+dz ]$ is
\begin{eqnarray}
   dN(z)=K \Sigma_{grb}(z)dz \int_{L_{min}(z)}^\infty \phi(L)dL, \label{sel}
\end{eqnarray}
where $K$ is a constant determined by detection efficiency. $L_{min}(z)$ is the luminosity threshold given by \cite{cao11}
\begin{eqnarray}
   L_{min}(z)=4\pi{D_L(z)}^2 P _{\rm th}k(z),\label{lmins}
\end{eqnarray}
where $P _{\rm th}$ = $2\times10^{-8}~\rm erg~s^{-1}cm^{-2}$. 
The value of k correction varies from 3.4 to 2.1 as the redshift increases from 0 to 10.

So, the normalized observable GRBs number density rate is
\begin{eqnarray}
   n(z)=\frac{dN{(z)/dz}}{\int_{z=0}^{\infty}dN(z)} .
   \label{n_z}
\end{eqnarray}
The solid lines in Fig. \ref{fig1} show the predicted observable GRB number distribution in the redshift space with two different SFR rate fittings. The red line uses the SFR in
Eq. (\ref{SFR1}), and the blue line uses the SFR in Eq. (\ref{SFR2}). There are 743 \emph{Swift} LGRBs in total, but only 1/3 of them have redshifts. The deviation between observation and models in redshift [1, 2.5] may be caused by selection effects, for instance the redshift desert effect.
Both models generally follow the same distribution, SFR1 with a more pronounced peak at about $z=1$.
\subsection{Reconstructing the Redshift Distribution of BATSE and \emph{Fermi}/GBM GRBs Sample}
For BATSE and Fermi/GBM, we assume that the GRBs intrinsic number distribution and luminosity function in the redshift space is the same as \emph{Swift}. The only difference is the $L_{min}$ in Eq. (\ref{lmins}) because of a different instrument sensitivity. For BATSE, we choose $P _{\rm th}$ = $1\times10^{-7}~\rm erg~s^{-1}cm^{-2}$ in the range $[50,\ 2000]$ keV \cite{yon04}. For \emph{Fermi}, we choose $P _{\rm th}$ = $1.7\times10^{-7}~\rm erg~s^{-1}cm^{-2}$ in the range $[50,\ 300]$ keV (\emph{Fermi} website).
The $k(z)$ is calculated with the Band function. For BATSE, we choose the indices $\alpha=-1.1$, $\beta=-2.69$ and $E_p=228$ keV \cite{ada13}. For GBM, we choose the indices $\alpha=-1.32$, $\beta=-2.24$ and $E_p=261$ keV \cite{gru14}.

\begin{figure}
\begin{center}\includegraphics[angle=0,scale=0.65]{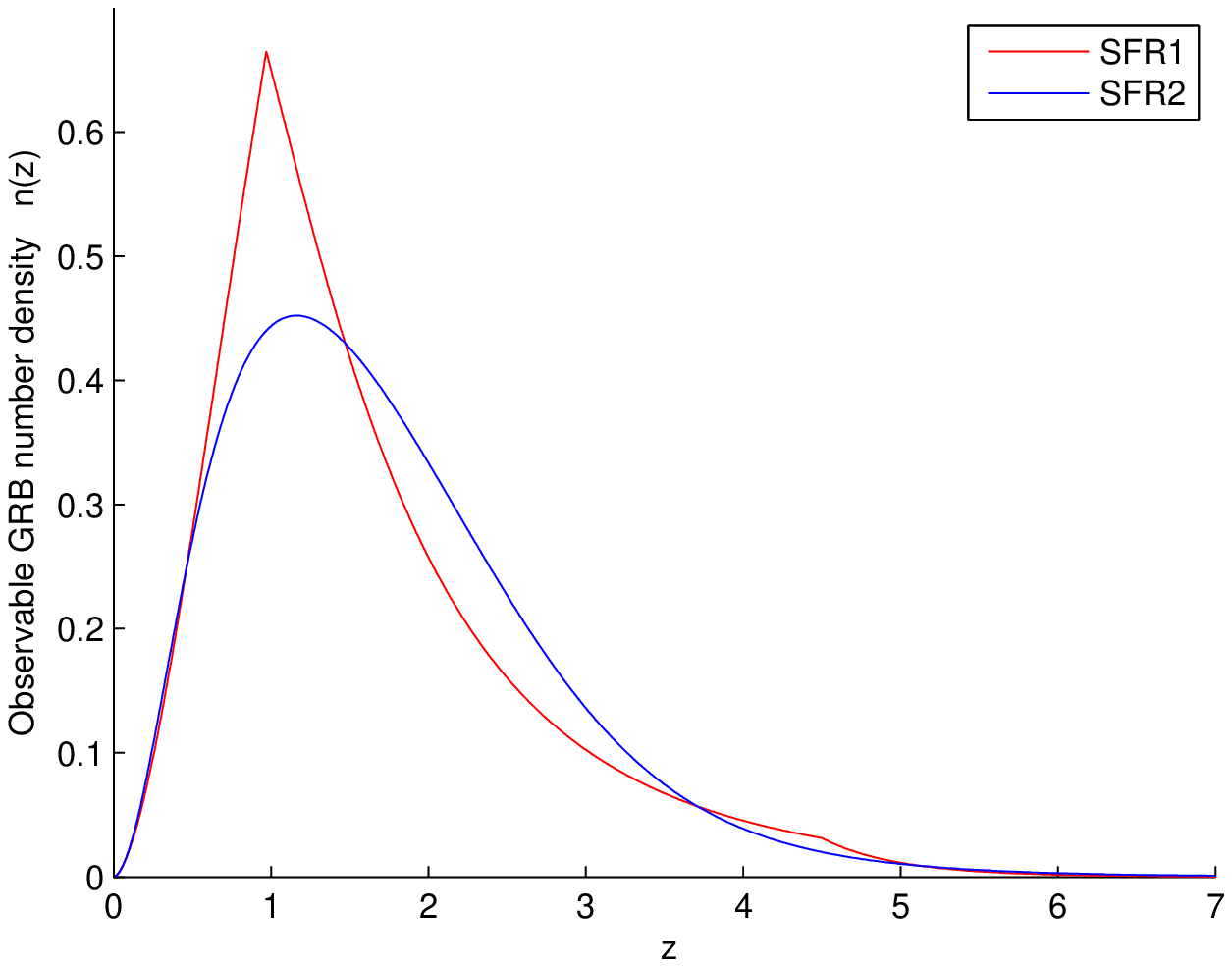}\end{center}
\caption{The normalized observable GRB number distribution in the redshift space for BATSE. The red line uses the SFR in
 Eq. (\ref{SFR1}), and the blue line uses the SFR in Eq. (\ref{SFR2}).
\label{fig1_2}}
\end{figure}

\begin{figure}
\begin{center}\includegraphics[angle=0,scale=0.65]{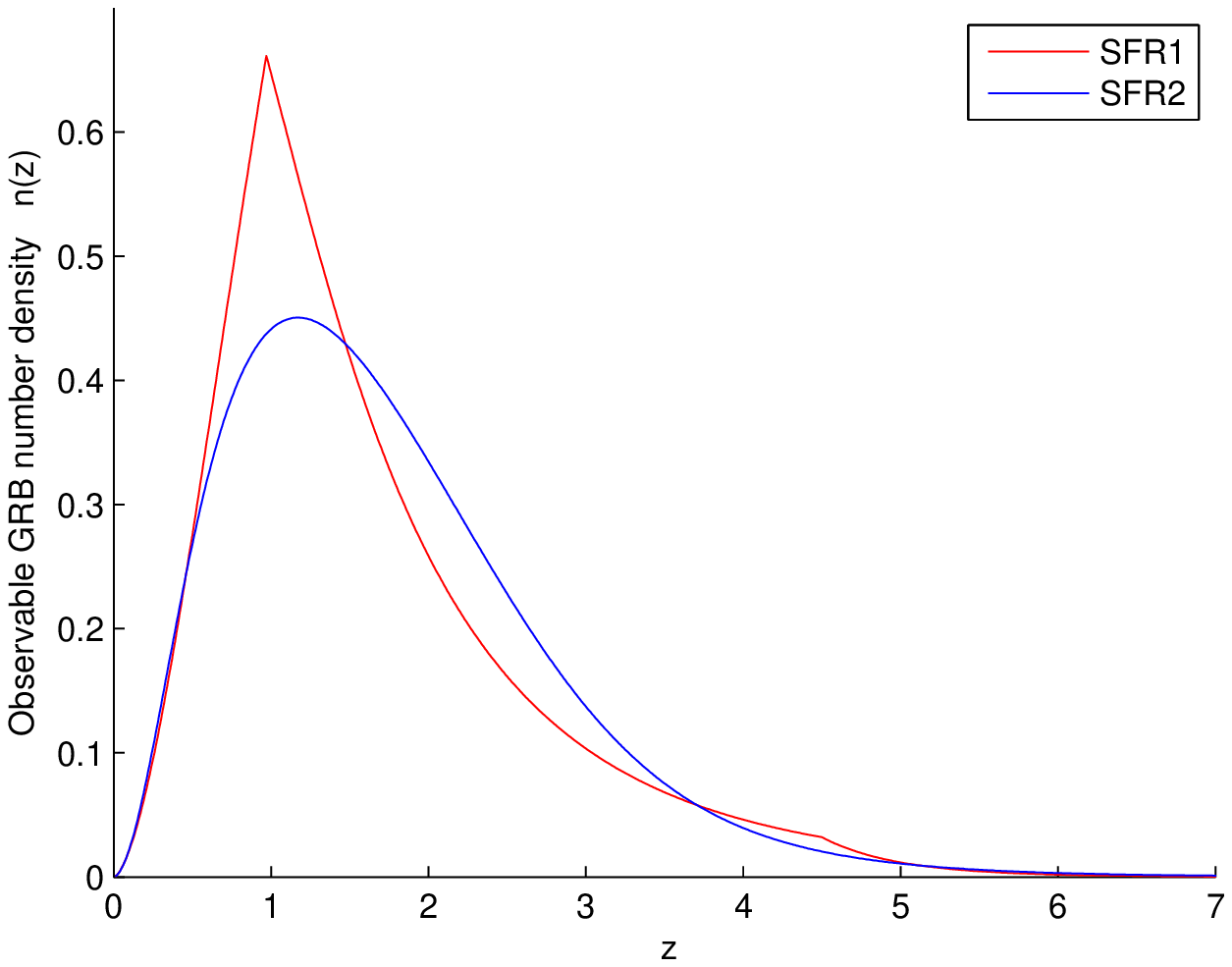}\end{center}
\caption{The normalized observable GRB number distribution in the redshift space for \emph{Fermi}. The red line uses the SFR in
 Eq. (\ref{SFR1}), and the blue line uses the SFR in Eq. (\ref{SFR2}).
\label{fig1_3}}
\end{figure}

Fig. \ref{fig1_2} shows the predicted observable GRB number distribution of BATSE in the redshift space with two different SFR rate fittings. The red line uses the SFR in
Eq. (\ref{SFR1}), and the blue line uses the SFR in Eq. (\ref{SFR2}).
Fig. \ref{fig1_3} shows the predicted observable GRB number distribution of \emph{Fermi}/GBM in the redshift space with two different SFR rate fittings. The red line uses the SFR in
Eq. (\ref{SFR1}), and the blue line uses the SFR in Eq. (\ref{SFR2}).

For \emph{Swift}, GBM and BATSE, we can see the normalized observable GRB number distributions in the redshift space are similar, so the thresholds has a limited influence on the normalized distribution.
\subsection{Calculating Magnification Bias}
\begin{figure}
\begin{center}\includegraphics[angle=0,scale=0.7]{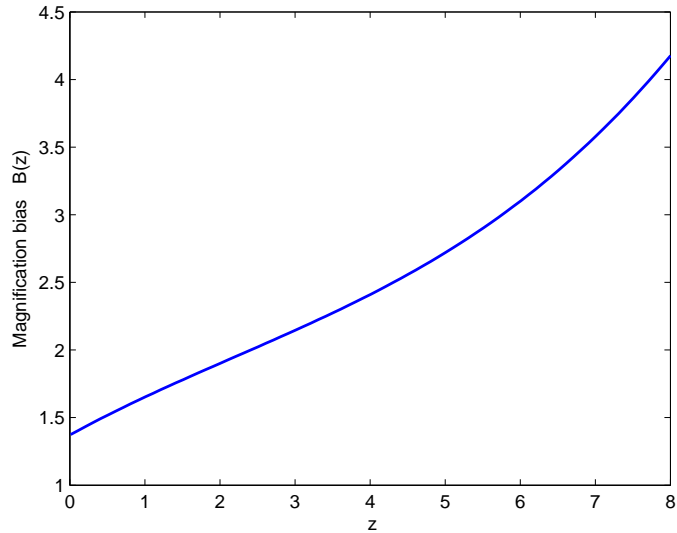}\end{center}
\caption{The magnification bias as a function of the source redshift $z_S$. The bias at lower redshift is determined by the lower luminosity power-law index $v_1$, and at higher redshfit determined by the higher luminosity power-law index $v_2$.
\label{fig2}}
\end{figure}

The magnification bias represents how lensed objects at redshift $z_S$ are overrepresented in any particular observed sample \citep{tur84}.
For the luminosity function (\ref{lum}) with two power-laws, $B(z)$ is the weighted average of the bias for each power-law
\begin{eqnarray}
B(z)=B_1 f_1(z) + B_2 f_2 (z),
\end{eqnarray}
where $f_1$ and $f_2$ are number fraction of GRBs with $L < L_b$ and $L < L_b$, and $B_1$, $B_2$ are the bias for each power-law function. According to \cite{tur84,sch92}, the bias for a single power-law can be simplified and we obtain $B_1= 1.3$ and $B_2=5$.
Fig. \ref{fig2} shows $B$ varies from $B_1$ to $B_2$ as the redshift increases, since low luminosity GRBs become undetectable at higher redshift.

Given $B(z)$, $n(z)$ and $P(z)$ for each detector, the expected sample size to produce one lensing case $N_0$ for SFR1 and SFR2 model are listed in Table 1.
\subsection{Influence of the Sampling Efficiency}
We define the sampling efficiency here for transient sources observation to be the ratio of the well recorded sample number to the whole events number that satisfy the observation sensitivity in the whole sampling field and time. For a telescope with an all sky FoV or a limited FoV but sampling in a constant field and running all the time before the whole sampling program end, the sampling efficiency is $f=1$. Many causes can decrease the sampling efficiency, for instance, a telescope with a limited FoV moving among different fields, or a ground based telescope with discontinuous observation for a certain field of sky because of the rotation of Earth, or incomplete data recording because of technical problems.

Compared to the lensing observation of quasars and galaxies, transient sources are quite sensitive to the sampling efficiency because of the short burst period \citep{li14}.  Ignoring technical recording problems, sampling efficiency for a certain field $s$ is given by
\begin{eqnarray}
  f(s)\equiv \frac{t}{T_0},
\end{eqnarray}
where $t$ is the practical sampling time and $T_0$ is the period of the whole sampling program covering, including observation gaps. For example, a telescope keeps sampling at a small beam for 12 hours per day, then the sampling efficiency is $f = 1/2$. For lensing observation, considering both the pair should be sampled, the expected sample size in Eq.  \ref{N0} should be enlarged by $1/f$.

Assuming the whole sampling area $S$ is consist of groups of small field $s$, the sampling program period is $T$, sampling efficiency in each small field $s$ is  $f(s)$, the burst rate on all sky is $R$, the expected lensing sample size $N = N_0/f$, where $N_0$ is given by Eq. (\ref{N0}).
Then the probability for the telescope to catch the double image is
\begin{eqnarray}
  P_{doub}=\int_{S}\frac{f(s)RT}{S_0 N}ds,
\end{eqnarray}
where $S_0$ is the area of the whole sky. Let $P_{doub}=1$, the expected sampling time to observe one lensing case is given by
\begin{eqnarray}
  T= \frac{S_0 N_0}{R} \frac{1} {\int_{S} {f^2(s)}ds}.
\end{eqnarray}

If the whole sampling field has a constant sampling efficiency $f(s)=f$, then the equation can be simplified as
\begin{eqnarray}
    T = \frac{S_0 N_0}{S R f^2 }, \label{st01}
\end{eqnarray}

Moreover, if it is a space telescope cyclically sampling all sky, then we get $S=S_0$, the equation can be written as
\begin{eqnarray}
    T = \frac{N_0}{R f^2 }.   \label{st02} 
\end{eqnarray}

Since $R=R_0/f$, where $R_0$ is the observed burst rate, the equation can be written as
\begin{eqnarray}
    T = \frac{N_0}{R_0 f }.   \label{st03}
\end{eqnarray}
Given $N_0$, $R_0$ and $f$, we can calculate the expected observation period.

Of course, if the calculated $T$ is shorter than the typical time delay between two images $\Delta t$ (as shown in \cite{li14}), $T$ will be only dependent of $\Delta t$.
\subsection{Expected Period to observe one lensing case for BATSE, \emph{Swift}/BAT and \emph{Fermi}/GBM}
 The BATSE orbit is about $400$ km above the earth, thus the FoV is about $0.65$ of whole sky excluding Earth. During the $3,323$ day operating period, there are about $2,390$ effective days for GRB observation. In all the $2,702$ GRBs observed by BATSE, only about $2,140$ with complete data because recording and communicating problems \citep{ada13}. Thus the sampling efficiency $f_{BSE}=0.65\times \left(\frac {2390\times 2140}{3323\times 2702}\right)\approx 0.37$. Since we primarily consider LGRBs here, we choose GRBs with $T_{90}>2$ s. There are about $1,530$ LGRBs in the $2,140$ samples, so the observed LGRB burst rate is about $R_0^{BSE}=170$ per year and the all sky burst rate is about $R=460$  per year.

  The sampling field of \emph{Swift} is all sky, but BAT can cover only about $0.16$ of the sky, so sampling efficiency upper limit is $f_{BAT}\approx0.16$. The observed LGRB burst rate about $R_0^{BAT}=80$ per year and the all sky burst rate is $R^{BAT}>500$ per year.

 \emph{Fermi}/GBM is similar to BATSE and the FoV is about $0.6$ of the whole sky. Without considering bad data and technical gaps, the upper limit of sampling efficiency is $f_{GBM}=0.6$. Up to GRB140414, there are about $1,120$ GRBs with $T_{90}>2$ s, so the observed LGRB burst rate is about $R_0^{GBM}=190$ per year and the all sky burst rate is $R^{GBM}> 320$ per year.

Given $N_0=1,170$ , $1,260$ and $1,250$ for \emph{Swift}, BATSE and \emph{Fermi} respectively, we can obtain the expected period for each telescope with Eq. (\ref{st03}). The results are listed in Table 2. From Table 2 we can see that, both BATSE and Fermi/GBM have a practical opportunity to get one lensed GRB pair, while Swift/BAT does not seem to catch any pair during the whole operation.
\section{Search for Strong Gravitational Lensing Events in the BATSE GRB Data}
There are $2,702$ GRBs in BATSE Current GRB Catalog. The basic table contains the location information of all $2,702$ GRBs. The fluence and flux table contains four channel fluence and peak flux information in $64$ ms, $128$ ms and $1,024$ ms of $2,135$ GRBs and the duration table contains $T_{90}$ and $T_{50}$ information of $2,041$ GRBs. The BATSE 5B GRB Spectral Catalog contains the location, $T_{90}$ and spectrum fitting result with five models for each of the $2,145$ GRBs \citep{ada13}. The spectral models are the single power-law model (SPL), the Band function (BAND), the Comptonized model (COMP), the smoothly broken power-law (SBPL) and the $Log_{10}$ Gaussian model (GLOGE). These fits are performed using 14-channel data covering energy range from $20$ keV-$2$ MeV, usually $2$-second resolution CONT data. For each model, there are two spectrums, the Peak flux spectrum over a $2.05$-second time range at the peak flux of the burst and the fluence spectrum over the entire burst duration.

In this paper, we search for lensing GRBs in the BATSE sample based on the BATSE 5B GRB Spectral Catalog, since according to the calculation in the previous section, it is more likely to include lensing sample as shown in Table 2. In order not to miss the lensing case of short GRBs although the calculation in the previous section are for LGRBs, we do not eliminate them in our sample. We use the Fluence spectrum for all models.

We construct the candidate sample by first choosing GRB pairs with angular separation less than $4^\circ$. BATSE GRBs have an average location accuracy about $1.7^\circ$, mostly in $3^\circ$. So $4^\circ$ will cover most of the potential lensing pairs. We obtain $2,889$ candidate pairs, some GRB involving in more than one pair.

We then searched for lensing samples according to the following four selection criterions:

Firstly, the earlier, the brighter. According to gravitational lensing theory, the earlier observed image should be brighter than the later one. In order not to miss the equally bright pairs, here we set the brightness ratio defined by the earlier to the later to be $r_{lum}>0.9$. We use the average flux (fluence/ intergration time) to represent brightness.

Secondly, similarity in spectra. Ideal lensing pairs should have the same spectrum. Considering the uncertainty in the fitting, here we choose a rough constraint not to rule out potential candidates. For the single power-law model, the index should satisfy $\Delta \lambda < 0.2$. For the Band function the index $\alpha$ and  $\beta$ should satisfy $\Delta \alpha < 0.2$ and $\Delta \beta < 5$. For the Comptonized model the index $\Delta \alpha < 0.2$. For the Smoothly broken power-law $\Delta \lambda_1 < 0.2$ and $\Delta \lambda_2 < 5$, For the $Log_{10}$ Gaussian model the full width at half maximum $\Delta FWHM < 5$.

Thirdly, similarity in duration. We constrain the $T_{90}$ ratio to be less than $1.5$ for pairs with the longer $T_{90}>10s$ and $T_{90}$ ratio is less than 2 for pairs with the longer $T_{90}<10s$.

 Lastly, lightcurve similarity. We plot and compare the lightcurve of the candidate pairs and compare the lightcurve by eyes. For the especially interesting pairs, we compare their four channel lightcurves respectively and study the spectrum parameter in detail. The data to plot lightcurves can be downloaded from ftp://cossc.gsfc.nasa.gov/ with energy range from $25-60$ keV, $60-110$ keV, $110-325$ keV and $>325$ keV.

For particularly bright GRBs, the BAND and SBPL functions are typically an accurate description of the spectrum, while for weaker bursts the COMP or GLOGE function is more acceptable. Bursts that have signal significance on the order of the background fluctuations do not have a detectable distinctive break in their spectra
 so the power law is the more acceptable function. To avoid missing potential lensing candidates, we let the pair survive if any one of its five models parameter satisfy all the selection criterions\citep{ada13}.
\section{Result}
\begin{figure}
\begin{center}\includegraphics[angle=0,scale=0.65]{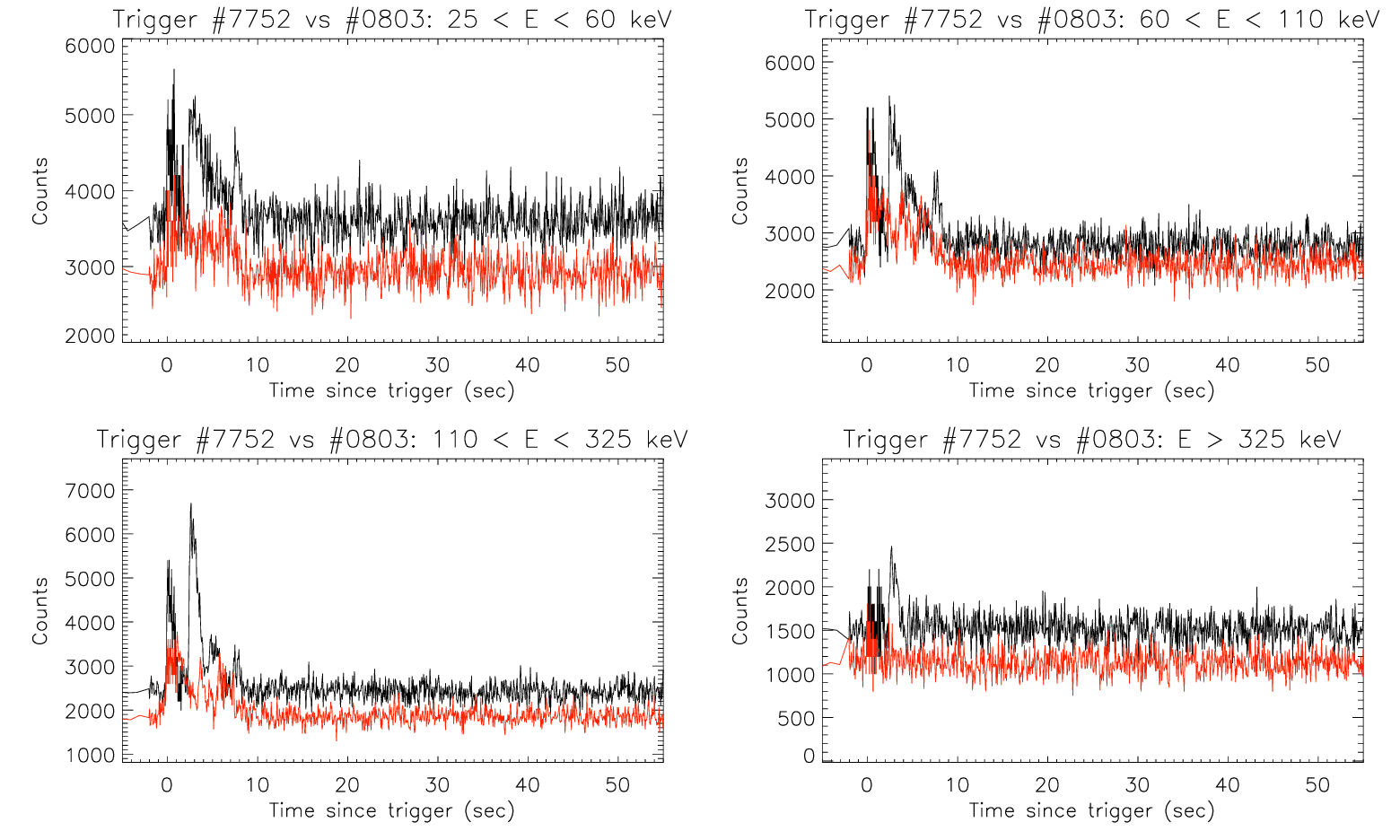}
\includegraphics[angle=0,scale=0.8]{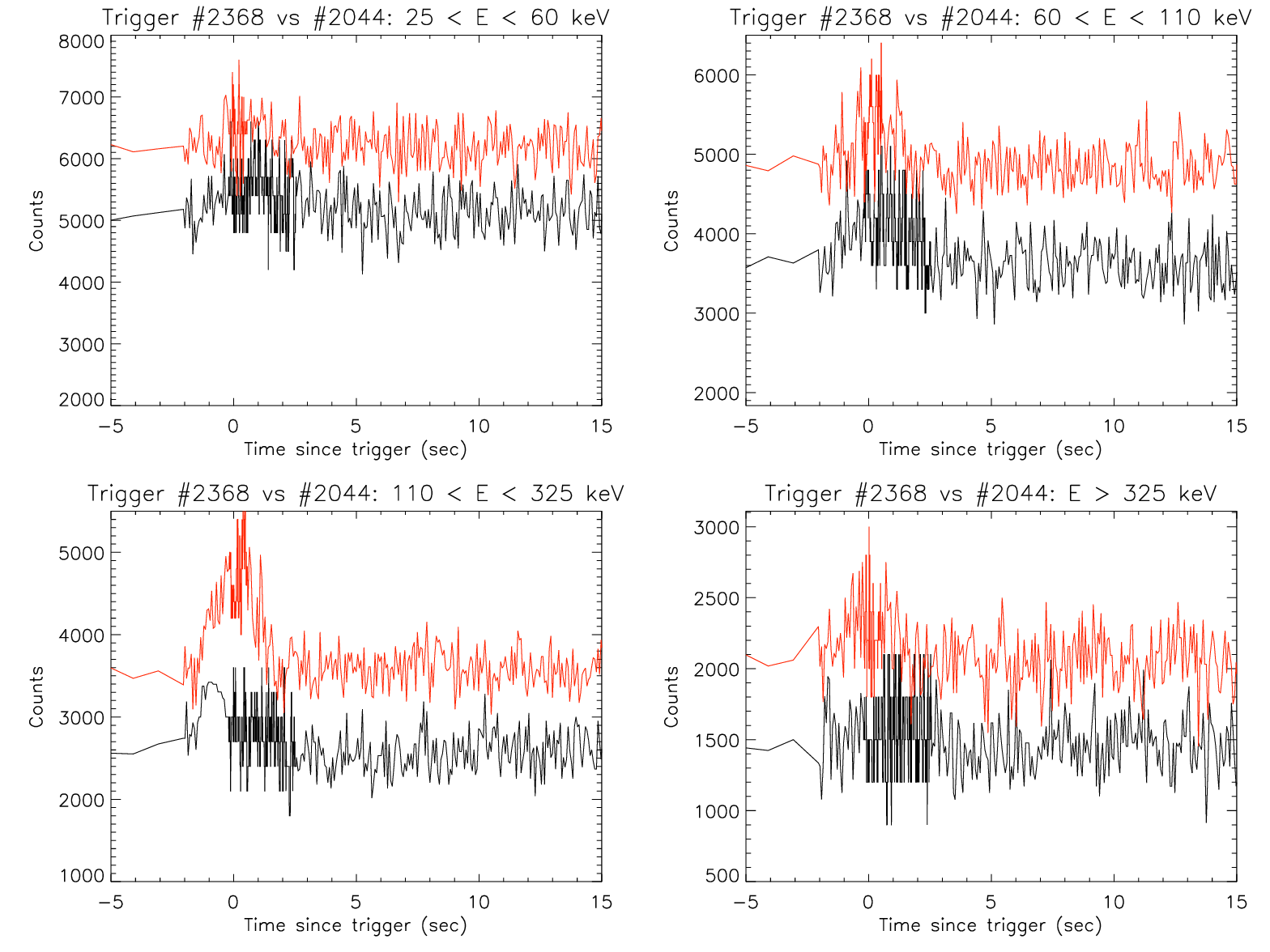}\end{center}
\caption{Four channel lightcurve comparison for 0803 vs 7752 and 2044 vs 2368.
\label{fig3}}
\end{figure}

\begin{figure}
\begin{center}\includegraphics[angle=0,scale=0.65]{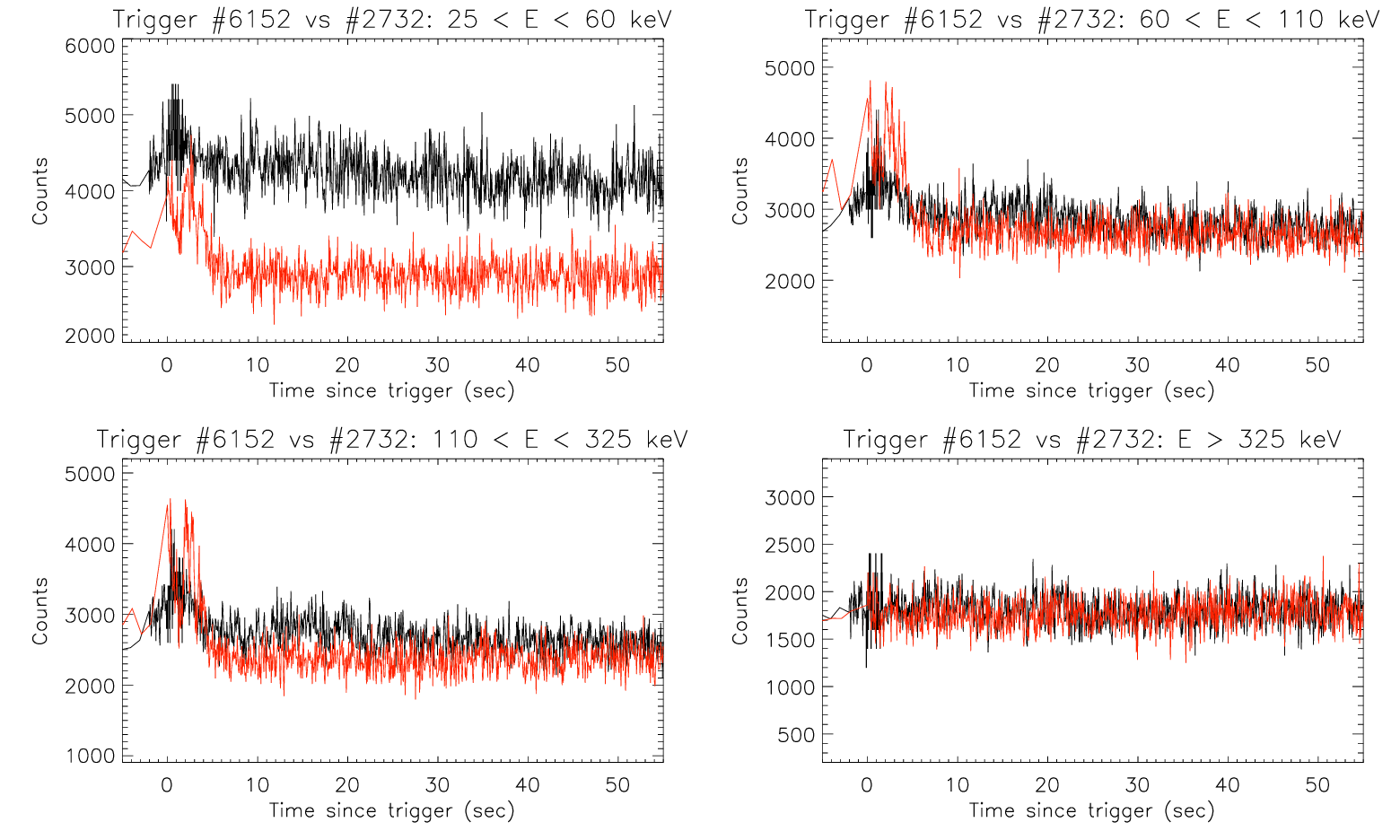}
\includegraphics[angle=0,scale=0.65]{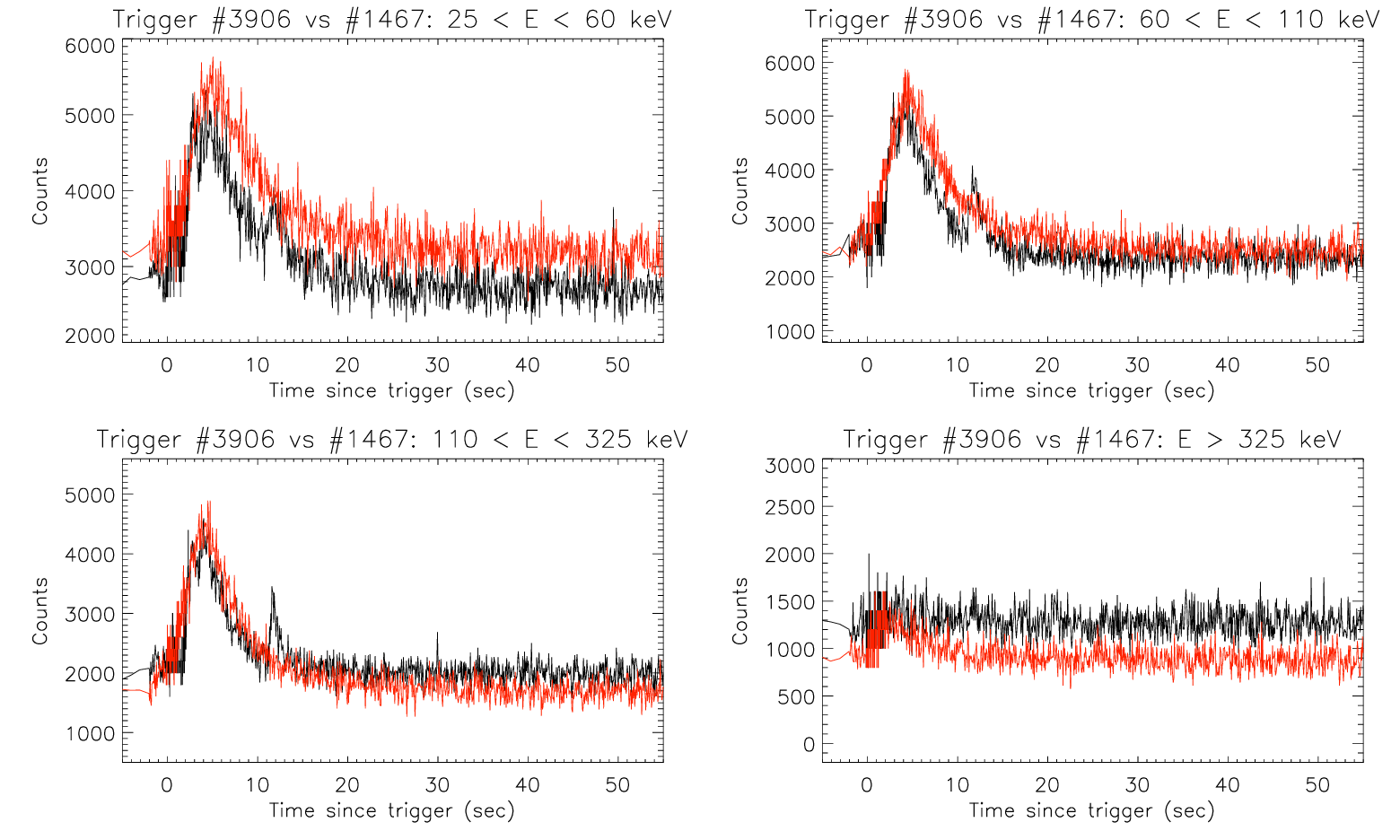}
\end{center}
\caption{Four channel lightcurve comparison for 2732 vs 6152 and 1467 vs 3906.
\label{fig4}}
\end{figure}

There are 57 pairs surving after the first three criterions. Then we plot the corresponding lightcurves in the range of $60-110$ keV. After lightcurve comparison, four pairs seem noteworthy. The trigger number are 0803 vs 7752, 2044 vs 2368, 2732 vs 6152, and 1467 vs 3906. The basic information of these bursts is shown in Table 3. Then we plot and compare all their four channels lightcurves, as shown in Fig. \ref{fig3} and Fig. \ref{fig4}. In each pair, the black line is for the first GRB, and redline is for the second GRB. For 2044 vs 2368, we enlarge the 2368 signal to make better comparison.

From Fig. \ref{fig3} and Fig. \ref{fig4} we can see that,
for 0803 vs 7752, the flux ratios in all channel are quite similar, but the peak profile appear slightly different.
For 2044 vs 2368, the flux ratios in four channel seem similar. We then checked the detail information of the pair in the Basic table, Fluence and Flux table and Duration table of the BATSE Current Gamma-Ray Burst Catalog. We list the information in Table 4. The location error for 2368 and 2044 are $6.06 ^\circ$ and $2.88^\circ$. Compared with the angular separation $\Delta \theta =3.88^\circ$, the location is consistent with an intrinsic arcsecond-scale separation. But the corresponding fluence ratios are not consistent with each other, thus we eliminate it.
For 2732 vs 6152, the flux ratio in each channel is not consistent with each other, so this is also eliminated.
For 1467 vs 3906, the profile and flux ratio are similar in each channel, but the $1467$ has an significant sub-peak about $5\sigma$ of the background fluctuation, which is hard to interpret with lensing effects.
\section{Discussion}
We have calculated the lensing probability in GRB samples by dark matter halos in foreground galaxies. The same method is applicable to other transient sources. From our calculation, the current BATSE GRBs sample is approaching the expected sample size to produce one observable lens case, thus we did a search for lensing GRB pairs in the BATSE current GRB catalog. We were unable to identify any lensing case. Since the expected period for BATSE to observe a lensing GRB pair is about $20$ years as shown in Table 2, which is twice of the BATSE operation period (9 years), the null result is consistent with our calculation.

Compared to BATSE, \emph{Swift}/BAT is with a higher sensitivity but much lower probability to detect lensing pairs.
Though BAT' FoV is only $1/4$ of BATSE, it is not the primary cause. If Swift is set to observe a certain part of sky all the time, from Eq. (\ref{st01}), we can see the expected period will drop from $94$ years to $15$ years, which would be promising because it is designed to be in use for about two decades.

Also, we note that the GRB observation for \emph{Fermi} is approaching to the lensing expected sample size. Since it may be in service for more than $15$ years, researchers have reason to anticipate the first GRB lens sample in future observations.

\section*{Acknowledgments}

The authors are grateful for the assistance from Dr. Huihua, Zhang. This work was supported by the National Basic Research Program (973
Program) of China (Grant No. 2014CB845800) and the NSFC grants
(No. 11373012).

\clearpage

\clearpage

\clearpage
~
\vspace{3cm}
\begin{table}[h]
\centering
\caption{The expected sample size to have one gravitational lensed GRB pair for different detectors, without considering the sampling efficiency. \label{tab0}}
\vspace{0.5cm}
\begin{tabular}{lcc}
\hline
\hline
~~~~~Model & SFR1 & SFR2\\
\hline
Sample size for BAT $N_0$~~~~ & ~$1170$~ & ~$1040$~\\
Sample size for BATSE $N_0$~~~~& ~$1260$~ & ~$1120$~\\
Sample size for GBM $N_0$~~~~& ~$1250$~ & ~$1110$~ \\
\hline
\end{tabular}
\end{table}

~
\vspace{3cm}
\begin{table}[h]
\centering
\caption{The expected period to observe one lensing case for different GRB detectors. \label{tab1}}
\vspace{0.5cm}
\begin{tabular}{lccc}
\hline
\hline
~Detector & BATSE ~~& ~\emph{Swift}/BAT ~&~ \emph{Fermi}/GBM \\
\hline
~Expected complete sample size $N_0$ & ~$1260$~ & ~$1170$~ & ~$1250$~\\
~Observed Burst Rate $R_0$($yr^{-1}$)& ~$170$~ & ~$80$~ & ~$190$~ \\
~Sampling efficiency $f$& ~$~0.37$~ & ~$~<0.16$~ & ~$~<0.6$~ \\
~Expected period $T_0$(yr) & ~$20$~ & ~$>94$~ & ~$>11$~ \\
~Observation time $T$(yr)~~~~ & ~ $9$~ & ~$10$~ & ~$6$~\\
\hline
\end{tabular}

\end{table}

\clearpage
\vspace{4cm}
\begin{table}[h]
\centering
\caption{The basic information of the four candidate GRB pairs. The Model column shows the pairs is accepted by which spectrum models. Numbers $1, 2, 3, 4, 5$ represent the BAND, COMP, GLOGE, SPL and SBPL, respectively. Values $\Delta \theta\ $, $\Delta T\ $ and $r_{lum}$ show the angular separation, time delay and the average flux ratio during the fluence integration period. If the pair is accepted by more than one model, then $r_{lum}$ is the average of the values of all models, which are usually equivalent.
\label{tab2}}
\vspace{0.5cm}
\begin{tabular}{lccccc}
\hline
\hline
~~~Pair & Model~~&~~$\Delta \theta\  (^\circ)$~~& $~~\Delta T\ $ (day) & $r_{lum}$& ${T_{90}}\ (s)$\\
\hline
0803 vs 7752 &~ 4~&~$3.93$~ & ~$2908$~ & ~$0.91$~ &~$7.23/8.64$~\\
2044 vs 2368 &~ 4~&~$3.88$~ & ~$205$~ & ~$1.44$~ & ~ $6.40/4.13$ ~\\
2732 vs 6152 &~ 3,5~&~$3.27$~ & ~$1186$~ & ~$2.14$~ &~$24.86/34.24$ ~\\
1467 vs 3906 &~ 2,3,4~&~$0.51$~ & ~$1346$~ & ~$1.10$~ &~$26.82/20.74$ ~\\
\hline
\end{tabular}

\end{table}

\vspace{5cm}
\begin{table}[h]
\centering
\caption{The detailed information of the the pair 2044 vs 2368 from the Basic table, Fluence and flux table and Duration table, T50 and T90 in unit of second, and fluence in unit of erg cm$^{-2}$.
\label{tab3}}
\vspace{0.5cm}
\begin{tabular}{lcccccccccc}
\hline
\hline
trigger&	RA	&DEC	&errRA	&T50	&errT50	&starT50	&T90	&errT90	&starT90\\
\hline
2368	&217.98&	-32.34	&6.06	&1.815	&0.17	&-0.768	&4.135	&0.954	&-1.856\\
2044	&213.46	&-33.07	&2.88	&1.344	&0.326	&-0.32	&6.4	&4.487	&-1.216\\
\hline
trigger&	name&	Fluence1&	err1&	Fluence2&err2&	Fluence3&	err3&	Fluence4&	err4\\
\hline
2368	&930602B	&3.83E-08	&1.41E-08	&6.82E-08	&1.55E-08&	2.27E-07	&3.95E-08	&8.46E-07	&5.58E-07\\
2044	&921109	&3.40E-08	&1.70E-08&	3.35E-08&	1.20E-08	&3.18E-07	&2.90E-08	&3.15E-06	&5.32E-07\\
\hline
\end{tabular}
\end{table}
\end{document}